\documentclass[twocolumn,twoside,slac_two]{revtex4}
\usepackage{graphicx}
\usepackage{fancyhdr}
\usepackage{hyperref}
\hypersetup{
    colorlinks=true,
    urlcolor=magenta,
}
\begin{document}

\title{Measurement of the muon-induced neutron seasonal modulation with LVD}

%

\author{N. Yu. Agafonova on behalf of the LVD Collaboration}
\affiliation{Institute for Nuclear Research RAS, prospekt 60-letiya Oktyabrya 7a, Moscow, Russia}

\begin{abstract}
Cosmic ray muons with the average energy of 280 GeV and neutrons produced by muons are detected 
with the Large Volume Detector at LNGS. We present an analysis of the seasonal variation of the neutron 
flux on the basis of the data obtained during 15 years. The measurement of the seasonal variation of the 
specific number of neutrons generated by muons allows to obtaine the variation magnitude of 
of the average energy of the muon flux at the depth of the LVD location. The source of the seasonal variation 
of the total neutron flux is a change of the intensity and the average energy of the muon flux. 
\end{abstract}

\maketitle

\thispagestyle{fancy}


\section{INTRODUCTION}

As it is known, there is seasonal variation of the muon flux at the sea level and underground \cite{var}. 
The variation is caused by the temperature and barometric effects associated with the increase of 
the atmosphere altitude in summer and decrease of it in winter. Temperature effect infuences the muon production 
in the $\pi$/K- decays, barometric effect influences the muon decays. For high-energy muons, which we 
detect at 3650 m w.e. underground there is a positive temperature effect. It is associated with the 
fact that far underground penetrate, mainly, muons from pions of the first generation, the number of which increases 
when the atmosphere density of the upper layers (at the altitude $\sim$ 20 km) falls due to expansion of the atmosphere.

Measurements of the muon intensity variation deep underground were performed by experiments 
MACRO \cite{Macro}, MINOS \cite{Minos}, AMANDA \cite{Amanda}, LVD \cite{Selvi}, IceCube \cite{Ice}, 
Borexino \cite{Borexino}.

In the LVD papers \cite{Selvi}, \cite{VKKL} the seasonal modulation of the muon intensity and the variation 
of the neutrons produced by muons in the LVD material were presented. The variation amplitude 
of the muon flux intensity $I_{\mu} = 1.5\%$ and phase ${\phi} = 185{\pm}15$ 
were measured during 8 years of observations. 
Moreover, it was obtained that the amplitude of the neutron number variation is higher than the muon 
intensity modulation amplitude almost 10 times as much in Ref.\cite{Mal}. The purpose of this work is to determine more accurately 
the magnitude of amplitude and phase of the muon-induced neutron seasonal modulation.

\section{LVD DESCRIPTION}
The LVD (Large Volume Detector) is located in underground laboratory in Gran Sasso (Italy) 
at the average depth of $\langle H \rangle$ = 3650 m w.e. \cite{Bari}. The LVD is multipurpose experiment, 
whose the main goal is the search 
for neutrino flux from the stellar core gravitational collapse. According to this program, the detector 
works since 1992 \cite{LVD}. The LVD consists of 3 towers containing 840 scintillation 1.5 m$^3$ 
counters forming 7 levels and 5 column.

Muons passing through the LVD generate hadronic and electromagnetic showers, which include 
 neutrons produced in $\pi A$ (hadronic showers) and $\gamma A$ (electromagnetic showers) interactions. 
The neutrons are slowed down in a scintillation counters and captured by protons 
($n + p \to d + \gamma$, $E_{\gamma} = 2.2$ MeV) 
or by iron nuclei ($n + Fe \to Fe + \gamma$, where the summed average energy ${\Sigma} E_{\gamma}$ is about 8 MeV). 
The neutron lifetime in scintillator is about 200 microseconds. The trigger for the neutron detection is the energy 
release in a counter more than 50 MeV. 
Trigger opens the time gate of 1 ms for detection of $\gamma$-quanta at energy threshold of 0.5 MeV.

\section{DATA ANALYSIS}
We analyzed the whole available data set with the detector in its final configuration, starting in 1 
January 2001 and ending in October 2016.

We have chosen good inner counters in all three towers. We have selected from muon data counters hitting by 
 muon trigger ($E_{tr} > 50$ MeV, $0 < t_{tr} < 250$ ns).
In hitting counters we search for gamma-quanta pulses at the energy from 1.5 to 12 MeV within time window 
50 - 350 $\mu$s. These pulses ($N_{tot}$) are produced by neutrons and background.

In the analysis of neutron seasonal variation, we used the value 
$N_n/N_{tr}$ (specific number of muon-induced neutrons), where $N_n$ is the number of neutrons detected in a 
counter and $N_{tr}$ is the number of triggers (number of muons crossing a counter). 
Using this value we can determine the variation amplitude without taken into account the LVD acceptance, 
the neutron detection efficiency, and data gaps for short-term shutdown of counters.


\begin{figure*}[!t]
\includegraphics[width=160mm]{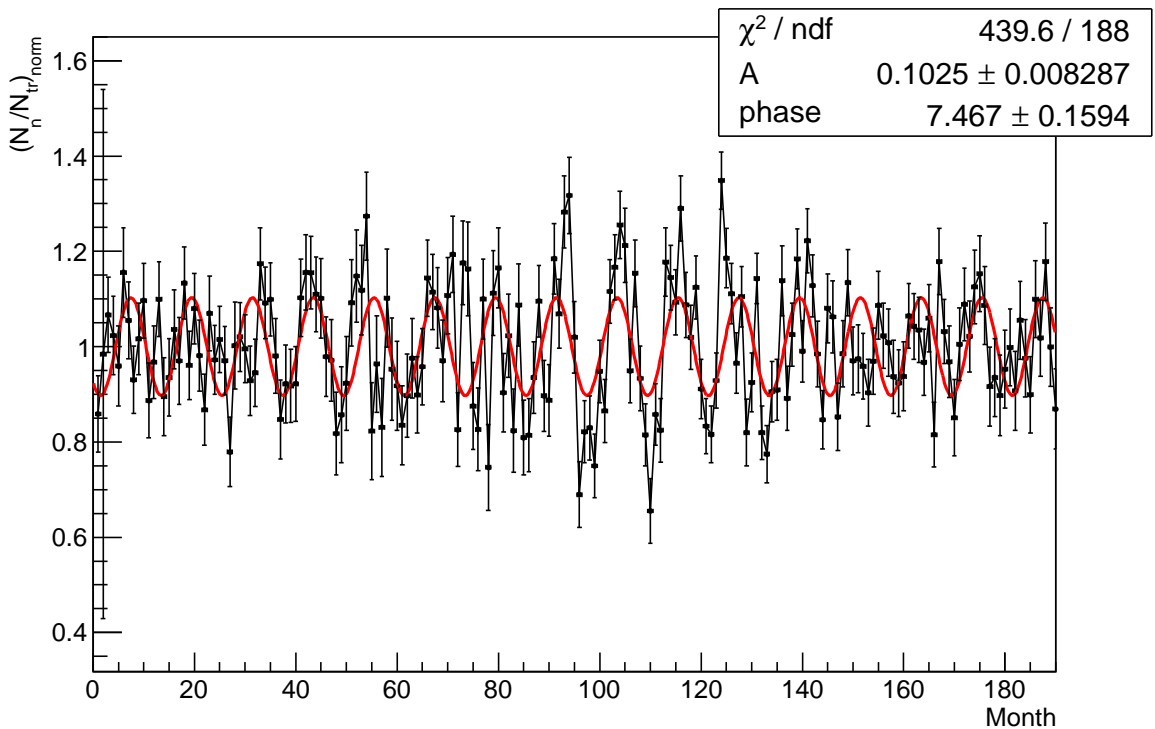}
\caption{The temporal behaviour of the value $N_n/N_{tr}$. The curve is the best fit with function
$f(t) = A \times cos(2{\pi}(t - {\phi})/T)+1$. Parameters $A$ and $\phi$ are free, $T$ = 12 months.}
\label{fig2}
\end{figure*}

\section{Determination of the variation amplitude of the number of the muon-induced neutrons}

We used two methods to determine the variation amplitude. These are the Epoch 
folding method and the Residual method.

\subsection{Epoch folding method}
Muon events in the inner LVD counters during 15 years (190 month) of observations were analyzed.
We determined the number $N_{tr}$ for a counter during each month and the number $N_n$ 
of neutrons within 50 - 350 $\mu$s.

To avoid background we assume that all pulses $N_{bg}$ within time interval of 450 - 750 $\mu$s are due to 
background and number of background pulses within interval of 50 - 350 $\mu$s is equal to the $N_{bg}$  number.

The specific number of muon-induced neutrons $N_n/N_{tr} = (N_{tot} - N_{bg})/N_{tr}$, where 
$N_{tot}=N_n + N_{bg}$, was determined for each month. In Figure~\ref{fig2} the $N_n/N_{tr}$ values are normalized 
per average year value $<N_n/N_{tr}>$. 

\begin{figure}[t]
\includegraphics[width=80mm]{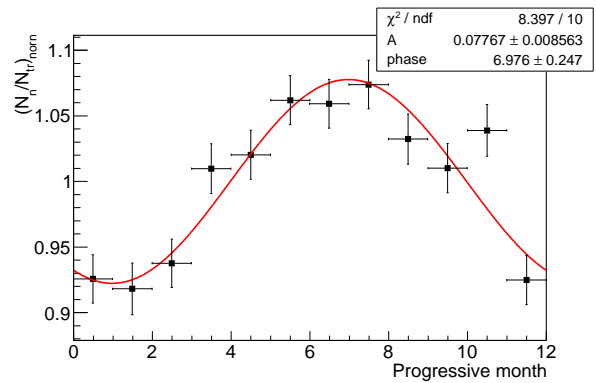}
\caption{Superposition of the mean monthly $(N_n/N_{tr})_{norm}$ values. The fitting curve  
 is function  $f(t) = 1 + A \times cos(2{\pi}(t-{\phi})/T)$. Parameter $T$ = 12 months is fix. 
Parameters $A$ and $\phi$ are free.}
\label{fig3}
\end{figure}

The Epoch folding method allows to obtain the amplitude modulation $\delta N_n/N_n$ and phase $\phi$ with 
sufficient accuracy. To do this annual data for the fifteen years of observations are superposed on one another.
The result of the procedure is presented in Figure~\ref{fig3}. 
When fitting the histogram on Figure~\ref{fig3} by the equation $N(t) = 1 + \delta N_n/N_n \times cos(2{\pi}(t - \phi )/T)$ 
for fixed $T$ = 12 months it was obtained $\delta N_n/N_n  = 7.7 \pm  0.8 \%$. 
The resulting phase is $\phi = 7.0 \pm 0.4 (stat) \pm 0.5 (sys)$ month.

\subsection{Residual method}
Large statistics allows to determine with good accuracy the number of muon-induced neutrons, which are produced 
during summer and winter months. To do this, we use the time distribution of gamma-quantum pulses in the 
energy range $1 - 12$ MeV in the time interval of $50 - 550 \mu$s after the muon trigger for 45 summer 
months (June, July, August) and 45 winter months (December, January, February).

\begin{figure}[t]
\includegraphics[width=70mm]{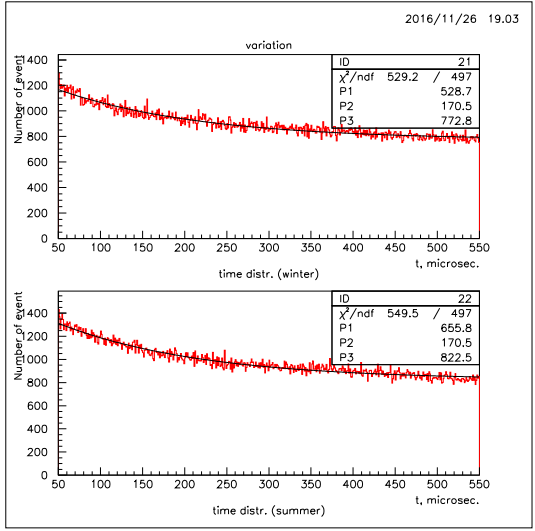}
\caption{Time distribution of the gamma-quantum pulses within $50 - 550 \mu$s for winter (upper panel) 
and summer (lower panel) months. The fitting curve is $f(t) = N_0 \times exp(-t/{\tau}) + B$.} \label{fig4}
\end{figure}

Time distributions were approximated using law $N_n(t) = N_0 \times exp(-t/{\tau}) + B$, where $\tau$  = 170 $\mu$s is 
exponent of gamma-quantum distribution, $B$ is a constant depending on the background 
conditions of counters, $N_0 \times \tau$ is the total number neutrons (Figure~\ref{fig4}).
It was found that the specific number of neutrons per a counter for summer months is
$N_n/N_{tr}^s =  111814/18695762 = 5.98 \times 10^{-3}$ , and 
$N_n/N_{tr}^w =  90143/17597826 = 5.12 \times 10^{-3}$ for winter months. 

The variation amplitude ($\delta N_n/N_n$) was defined as half a difference between the neutron specific number in 
summer (s) and winter (w), divided by the average value: 
$\delta N_n/N_n = 0.5 (N_n^s/N_{tr}^s - N_n^w/N_{tr}^w)/0.5 (N_n^s/N_{tr}^s + N_n^w/N_{tr}^w$). 
The value $\delta N_n/N_n = 0.077 \pm 0.002$ (stat.) $\pm$ 0.016 (sys.) was obtained.

\section{The modulation amplitude ${\delta} E$ of the average muon energy at the depth of 3650 m w.e.}
The number of neutrons generated by muon flux described by dependence 
$N_n \propto E_{\mu}^{0.78}$ \cite{UF}, where $E_{\mu}$ is the average energy of the muon flux.
Thus, the neutron number increasing in summer can be expressed as: 
$(N_n + \delta N_n)/N_n = ((E_{\mu} + \delta E_{\mu})/E_{\mu})^{0.78}.$

Substituting the measured value $\delta N_n/N_n = 0.077$ (the value determined by two methods), 
we obtained $(\delta E_{\mu}/E_{\mu}) (1 + 0.077)^{1/0.78}  - 1 \approx  0.10$. 
So, the average energy of the muon flux at the LVD depth increases by 10$\%$ in summer.

Consequently, the flux of muon-induced neutrons undergoes seasonal variation under 
the influence of two factors: 
a) change of muon intensity, b) change of muon average energy. 
The factor b) is dominant. Thus, the modulation amplitude  $\delta {\Phi}_n$ of the total neutron flux ${\Phi}_n$ is
$ \delta {\Phi}_n = (1 + \delta I_{\mu}) \times (1 + \delta N_n) -1 = 1.015 \times 1.077 - 1 = 0.093$, 
i.e. $\delta {\Phi}_n(I_{\mu}, N_n) = 9.3\%$.

\section{CONCLUSION}

The seasonal variation of muon-induced neutrons per muon was found on the basis of data for 15 years. 

The measured characteristics of the neutron variation indicate seasonal variation in 
the average energy of muons at the LVD 
depth with amplitude of $10 \%$, i.e. $ E_{\mu} = 280 \pm 28$ GeV.

Previously it was assumed that the flux of cosmogenic neutrons is proportional to the amplitude of muon intensity 
variation $1.5 \%$. We have shown that the neutron flux has an amplitude of seasonal variation in 6 times more, 
because the average muon energy also varies with the amplitude of $\sim  10 \% $.

\begin{acknowledgments}
This work was supported by the Russian Foundation for Basic Research (project No. 15-02-01056 a) 
and the program of investigations of the Presidium of Russian Academy of Sciences 
High Energy Physics and Neutrino Astrophysics.
\end{acknowledgments}

\bigskip 

\end{document}